\documentclass[%
reprint,
superscriptaddress,
showpacs,preprintnumbers,
nofootinbib,
 amsmath,amssymb,
 aps,
prb,
]{revtex4-1}

\usepackage[cp1251]{inputenc}
\usepackage{graphicx}
\usepackage{dcolumn}
\usepackage{bm}

\begin{document}


\title{Doped Mott-Hubbard materials with a low quasiparticle transparency}


\author{V. A. Gavrichkov}
\affiliation{L. V. Kirensky Institute of Physics, Siberian Branch of Russian Academy of Sciences, 660036 Krasnoyarsk, Russia}


\date{\today}

\begin{abstract}
Based on the Wilson's criterion metal-insulator, extended to materials with strong electronic correlations, we have identified a specific class of the materials, which is not associated with their usual classification into Mott-Hubbard and charge transfer dielectrics. The local symmetry of these materials leads to a disappearance of quasiparticle states (so-called first removal or fist extra states) in the Hubbard's gap. It's especially unusual for the doped materials, where quasiparticles - charge carriers  can disappear or appear under external factors, but the Mott transition has not yet been achieved. In the work we introduced the so-called a "quasiparticle transparency", and also provided specific experiments to identify the materials with the low quasiparticle transparency. Some examples of these materials with a spin crossover under high pressure and showing Jahn-Teller nature have been observed.

\end{abstract}

\maketitle


\section{\label{sec:intr}Introduction\\}

The satisfiability of known metal-insulator(MI) criterion $W \mathord{\left/
 {\vphantom {W {U \sim \left( {R + 1} \right)}}} \right.
 \kern-\nulldelimiterspace} {U \sim \left( {R + 1} \right)}$
(where $R$ - orbital degeneracy) of the Mott-Hubbard transition  depends on the characteristic material scale, because a width $W$  of the quasiparticle band decreases with decreasing scale of materials, and the Coulomb interaction $U$ increases due to the weakening of screening effects. The MI transition itself can also be controlled by external effects, which change physical properties of the material (thermal expansion, pressure, optical pumping, etc.). Of particular interest is the transition of Mott-Hubbard insulators  to the metallic state induced by a doping effect, since the doped materials have unique properties, e.g. high temperature superconductivity in the 2D perovskite cuprates~\cite{Kohsaka_etal2008, Yang_etal2008} and colossal magnetoresistance in 3D manganites ~\cite{Saitoh_etal2000, Mannella_etal2005}. Moreover, in both materials the pseudogap effect is observed. However, it's hard to imagine that the initial MI criterion is valid in all  various cases.

To see the origins of the problem, it is enough to look at the "band structure" within a framework of the formalism of Hubbard operators \cite{Hubbard_1963} in the zero-hopping approximation. The formalism is necessary to detect the effects of many-electron local states in the number of quasiparticles in the band. In the zero-hopping approximation, according to the initial MI criterion, the material must be an insulator ($W=0$), but the MI criterion based on Wilson's ideas ~\cite{Wilson_1931} for the doped Mott-Hubbard materials shows non-trivial scale invariant results, independent on band width $W$.
The  purpose of  our   work is to construct and apply Wilson's MI criterion concerning  a  system  of itinerant electrons in  the  analytical form for  the doped Mott- Hubbard materials. The approach includes a key statement that if an electron system consists of  completely occupied and empty bands, it is an  insulator, otherwise, it is a metal, where however the  spectral density  of  quasiparticles depends on  the doped carrier concentration $x$ due to many-electron  effects.\\

\section{\label{sec:II} Wilson's MI criterion  for the doped materials}

To extend  the Wilson's MI criterion, we will further follow the work ~\cite{Gavrichkov2015} where it was demonstrated that doped 2D perovskite cuprates have metallic conductivity, and there are no forrbidden quasiparticle states. Although there are some features here (e.g. an impurity potential effect and associated states), we will consider the criterion taking into account the many-electron effects only. Our extended approach uses the fact that the optical intracell $dd$-transitions with their ($l$-orbital, $S$-spin)-selection rules in the transparency window and optical charge transfer transitions in the oxides can be observed at the same $3d$-states.~\cite{Krinchik_etal1969, Eremenko_etal1969}

In the first approximation we can assume that the quasiparticles are unit cell excitations which  can be represented graphically as the single-particle transitions between different sectors $N_h=...(N_{h0}-1), N_{h0}, (N_{h0}+1),...$ of the configuration space of unit cell ($N_{h0}$-hole number per cell in undoped material, see Fig.\ref{fig:1}).~\cite{Ovchinnikov_etal2012} Each of these transitions forms the $r$-th quasiparticle band, where the root vector $r=\{ii'\}$ in the configuration space numerates the initial $i$ and final $i'$ many-electron states in the transition. The quasiparticle transitions with increasing or decreasing the electron is formed the conduction or valence bands respectively. Here, it is convenient to start with Lehmann's representation for Green's function $G_{fg\sigma} ^{\lambda\lambda}$
of the intracell Hamiltonian $H_0$  with respect to the family of single-particle operators  $c_{f\lambda\sigma}^{(+)}$ and their matrix elements in the basis of ${\left| {({N_{h}},{M_S})_i} \right\rangle} $ - eigenstates of $H_0$  ($S$ and $M$- spin and spin projection of the multielectron cell eigenstate):
\begin{widetext}
\begin{equation}
\sum\limits_{\lambda \sigma } {G_{fg\sigma }^{\lambda \lambda } }  = \sum\limits_{\lambda \sigma } {\left\langle {\left\langle {{c_{f\lambda \sigma } }}
\mathrel{\left | {\vphantom {{c_{f\lambda \sigma } } {c_{g\lambda \sigma }^ +  }}}
\right. \kern-\nulldelimiterspace}
{{c_{g\lambda \sigma }^ +  }} \right\rangle } \right\rangle }  = \delta _{fg} \sum\limits_{rr'} {\sum\limits_{\lambda \sigma } {\gamma _{f\lambda \sigma } \left( r \right)\gamma _{f\lambda \sigma } \left( {r'} \right)D_{fg}^{rr'} \left( E \right)} = }
= \delta _{fg} \sum\limits_{rr'} {\delta _{rr'} \sum\limits_\sigma  {\frac{{F_r \left( x \right)\chi _{rr}^\sigma  }}{{E - \Omega _r }}} },
\label{eq:1}
\end{equation}
\end{widetext}
where matrix elements:
\begin{eqnarray}
 {\gamma _{\lambda \sigma }}\left( {{r}} \right)& = &\left\langle ({{N_{h} + 1,{{M'}_{S'}}})_\tau} \right|c_{f\lambda \sigma } \left|({{N_{h},{M_S}})_\mu}\right\rangle\times \nonumber  \\
 & \times &\delta \left( {S',S \pm |\sigma| } \right)\delta \left( {M', M + \sigma } \right),
\label{eq:2}
\end{eqnarray}
and
\begin{equation}
\chi _{rr'}^\sigma   = \sum\limits_{\lambda } {\gamma _{\lambda \sigma }^* \left( r \right)\gamma _{\lambda \sigma } \left( {r'} \right)}
\label{eq:3}
\end{equation}

\begin{figure}
\includegraphics{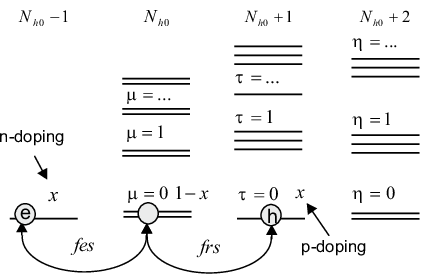}%
\caption{\label{fig:1} $E_i({N_{h}},{M_S})$ - energy level scheme of the configuration space based on cell states with hole numbers per cell $N_h=N_{h0}-1, N_{h0}, N_{h0}+1,..$, where $i=\mu, \tau, \eta$ and $N_{h0}$ is a hole number per cell in the undoped material. The $e$ - and $h$ - circles point to  the occupied ground cell states of electron and hole doped material respectively. A solid line with arrows corresponds to the $fes$ and $frs$ quasiparticles.}
\label{fig:1}
\end{figure}

for the $p$ - and $n$ - quasiparticle states in the valence and conduction bands respectively, where the total space of root vectors $\{r\}=...+\{r_{12}\}+\{r_{23}\}+..., (\{r_{12}\}=\{\mu\tau\},\{r_{23}\}=\{\tau\eta\}$ and so on, see Fig.\ref{fig:1}). An occupation factor $F_r\left( x \right)$ is the probability to detect a cell in any of the $i,i'$ states participating in the $r$-th transition, and
$\Omega^v _r=E_i({N_{h}},{M_S})-E_{i'}({N_{h}+1},{{M'}_{S'}})$ and $\Omega^c _r=E_i({N_{h}-1},{M_S})-E_{i'}({N_{h}},{{M'}_{S'}})$ are the quasiparticle energies in the $r$-th valence and conduction bands respectively. In the PM phase of doped material factor of ocuppation has a form:
\begin{equation}
F_{r_{12}}\left( x \right) = \frac{{1 - \alpha x}}{{2S + 1}},
\label{eq:4}
\end{equation}
where $\alpha=1-({2S+1})/({2S'+1})$ is proportional to the ratio of the spin multiplets of $i,i'$ states participating in the $r_{12}$ - (from the subspace $\{r_{12}\}$) transition between the ground states ${\left| {({N_{h0}},{M_S})_{i=0}} \right\rangle} $ and ${\left| {({N_{h0}+1},{M'_{S'}})_{i'=0}} \right\rangle}$ indicated by the arrow in Fig.\ref{fig:1}

The Green's function in Eq.(\ref{eq:1}) is yet free from the shortcomings of the hydrogen-like (s-)representation and low-energy approximations, because we do not restrict ourselves to choose the intracell Hamiltonian $H_0$, and are ready to work with all of the  ${\left| {({N_{h}},{M_S})_i} \right\rangle}$ states. Taking into account the specifics of the cuprates, we will consider the p - doped materials, where a valence states number is equal to the sum over all quasiparicle states:
\begin{widetext}
\begin{equation}
 N_v(x)= {\sum\limits_{\lambda\sigma}  {\sum\limits_{r}
{{\gamma _{\lambda \sigma }}^2\left( r \right)}
{{ \int\limits_{}  {dE\left( { - \tfrac{1}{\pi }}\right){\operatorname{Im} {D_0^{r}}\left( {E} \right)}}_{E + i0}}} }} =N_v^{12}(x)+N_v^{23}(x),
\label{eq:5}
\end{equation}
\end{widetext}
where $N_v^{12}(x)$ and $N_v^{23}(x)$ are the contributions from the quasiparticles with the root vectors $r$ from the $\{r_{12}\}$ and  $\{r_{23}\}$ subspaces because the other states of ${\left| {({N_{h}},{M_S})_i} \right\rangle} $ in p - doped material are not occupied,
and there is a zero probability $F_r\left( x \right) =0$ to detect a cell in these states at a low temperature.  Wilson's condition at the insulating state, which we are interested, it is
\begin{equation}
N_e-x=N_v(x),
\label{eq:6}
\end{equation}
 where $(N_e-x)$ is a total electron number per cell of hole doped material. That is, if the number of electrons in cell equals the number of valence quasiparticle states, the doped material is an insulator.

To obtain the Fermi level position in the degenerate doped material at zero temperature
we could carry out the integration on the right side of the equation
\begin{equation}
x={\sum\limits_{\lambda\sigma}  {\sum\limits_{r}
{{\gamma _{\lambda \sigma }}^2\left( r \right)}
{{ \int\limits_{E_F}  {dE\left( { - \tfrac{1}{\pi }}\right){\operatorname{Im} {D_0^{r}}\left( {E} \right)}}_{E + i0}}} }},
\label{eq:7}
\end{equation}
over the top valence band of the first removal electron states  ($frs$) with the lowest binding energy (see Fig.\ref{fig:1}), and this is sufficient at the actual concentrations $x \sim 0.1$  as a rule.  However, this is not sufficient, when the hole concentration $x$ exceeds the number of quasiparticle states in the top valence band $x \gg {N_{frs}}$, because the number of $frs$ quasiparticle states $N_{frs}$ may be very small. Therefore, the solution Eq.(\ref{eq:7}) has the features at $N_{frs}\rightarrow0$. To understand this, we will obtain the a total number of valence quasiparticle states $N_v$ as a function of both the doping concentration $x$, and $N_{frs}$:

\begin{equation}
N_v (x, N_{frs})= N_v^{12}+N_v^{23},
\label{eq:8}
\end{equation}
where the contributions $N_v^{12}$ and $N_v^{23}$ from the subspaces $\{r_{12}\}$ and $\{r_{23}\}$  are calculated in the Appendix~\ref{ap:1}, and the root vectors $r$ characterizes a specific quasiparticle band: if $r  = \left\{ {\nu _0 ,l_0 } \right\}$ or $r  = \left\{ {\tau _0 ,l_0 } \right\}$ in Eqs.(\ref{eq:2}) and (\ref{eq:3}), then we are dealing with $fes$ or $frs$ quasiparticles in Fig.\ref{fig:1}, respectively. By following this approach, we  obtained a MI criterion:

\begin{equation}
N_{v}(x) =N_e-x(1-N_{frs})
\label{eq:9}
\end{equation}
which is characterized by a condition: $N_{frs}=0$ ( - insulator) or $N_{frs}\neq 0$ ( - metal), and $N_{frs}$ is calculated in the Appendix~\ref{ap:1}. Indeed, at conditions $N_\lambda=1$ and $N_e=(1-x)$ we always obtain  a simple metal with $N_{frs}=2$ and $N_v(x)=(1+x)$ valence states,
as it takes place in the Hubbard model, where the highly spin (triplet) states are simply not available. The criterion is based only on the properties of completeness of a set of states ${\left| {({N_{h}},{M_S})_i} \right\rangle}$ in the configuration space of the cell, and the number of states $N_{frs}$ depends on their spin and orbital nature.

\section{\label{sec:III} Quasiparticles in  doped material with low quasiparticle transparency}

From Eq.(\ref{eq:9}) it follows that the doped material can show both the metallic $N_v(x)>(N_e-x)$, and dielectric properties $N_v(x)=(N_e-x)$ at the $N_{frs}>0$ or $N_{frs}=0$ respectively. The physical meaning of the MI criterion lies in the matrix value $\chi _{rr'}^\sigma $ (see Eq.(\ref{eq:3})) that in the Hubbard operators representation corresponds to the quasiparticle as a sequence of intracell transitions between multielectron cell states. If the single-particle transitions are forbidden by any symmetry, the charge carriers is missing. The doped particles (electron or holes) are in local multielectron states ${\left| {({N_{h}},{M_S})_i} \right\rangle}$, but there is no peak in the single-particle density of states, and the matrix value $\chi _{rr'}^\sigma $ defined in the root vectors space $\{r\}$ can be called "quasiparticle transparency" of the doped material.

\subsection{\label{sec:PIE} Pressure induced effects\\}
The $frs$ states can be prohibited at the $\delta (S',S\pm|\sigma|)=0$ in Eq.(\ref{eq:2}) and Eq.(\ref{eq:3}) (s-forbidden $frs$ quasiparticles).  However, can forbidden $frs$($fes$) states really exist in any doped Mott-Hubbard materials? A review of Tanabe-Sugano diagrams~\cite{Tanabe1956} shows that the low transparency effects in the materials with $3d$  elements in an octahedral environment are unlikely. Indeed, the ground states in different sectors of the configuration space are connected by non-zero matrix elements (\ref{eq:2}) of single-particle operators. However, the nature of the ground state of the transition element ion  depends on the applied pressure, while some materials with $3d^k$ ions  at $3 < k < 8$: the transition metal oxides, transition metal complexes, metal-organic molecules and molecular assemblies exhibit spin crossover, with increasing pressure from ambient pressure~\cite{Lyubutin2009, Gavrichkov2020, Halder2002, Ohkoshi2011, Wentzcovitch2009, Hsu2010, Liu2014, Sinmyo2017, Marbeuf2013, Nishino2007, Ovchinnikov2004, Saha2014}.
The spin crosover ocurrs due to competition between the energy crystal field $10Dq$ and the intra-atomic Hund exchange $J_H$, for example, in the interesting material FeBO$_3$~\cite{Knyazev2022}, where the Fe$^{3+}$ ion is in a high-spin configuration $3d^5 \left( {t_{2g}^3 e_g^2 } \right)$ at the ambient pressure. The energy of the high and low spin states in the $N_0(3d^5)$ sector can be presented in the form~\cite{Ovchinnikov2008}:
\begin{eqnarray}
E_{hs}&=&E_c(d^5)-10J_H \nonumber \\
E_{ls}&=&E_c(d^5)-20Dq-4J_H
\label{eq:10}
\end{eqnarray}
Eq.(\ref{eq:10}) shows that spin crossover $\left. S \right|_{P < P_S }  = {5 \mathord{\left/
 {\vphantom {5 {2 \leftrightarrow \left. S \right|_{P > P_S }  = {1 \mathord{\left/
 {\vphantom {1 2}} \right.
 \kern-\nulldelimiterspace} 2}}}} \right.
 \kern-\nulldelimiterspace} {2 \leftrightarrow \left. S \right|_{P > P_S }  = {1 \mathord{\left/
 {\vphantom {1 2}} \right.
 \kern-\nulldelimiterspace} 2}}}$
in the ground state is possible at a certain pressure $P_S$ ($=48 - 54$ GPa for the iron borate~\cite{Knyazev2022}) corresponding to the crystal field $10Dq=3J_H$. Here and below, $E_c$ is a part of the term energy independent of the Hund exchange $J_H$ and the crystal field $10Dq$~\cite{Ovchinnikov2013}. Similarly in sector $N_+$ for the $3d^6$ configuration
\begin{eqnarray}
E_{hs}&=&E_c(d^6)-4Dq-10J_H \nonumber \\
E_{ls}&=&E_c(d^6)-24Dq-6J_H
\label{eq:11}
\end{eqnarray}
for spins $S_{\nu_0}=2$ and $S_{\nu_0}=0$, respectively. This shows that crossover in the ground state of the term is possible,  at the condition $10Dq=2J_H$, in accordance with the pressure $P_{min}$. Similarly, in sector $N_-$ for the $3d^4$ configuration
\begin{eqnarray}
E_{hs}&=&E_c(d^4)-6Dq-6J_H \nonumber \\
E_{ls}&=&E_c(d^4)-16Dq-2J_H
\label{eq:12}
\end{eqnarray}

\begin{figure}
\includegraphics{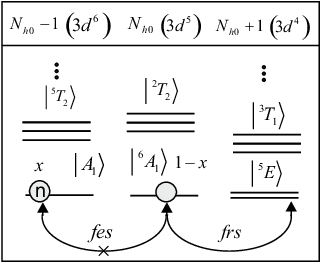}
\caption{Ground state crossover $S_{\nu_0}=2\leftrightarrow 0$ in the  $N_{h0}-1$ sector  FeBO$_3$ under the pressure $P_{min}<P<P_{S}$. A semiellipse with a cross shows forbidden $fes$ quasiparticles.}
\label{fig:2}
\end{figure}

Eq.(\ref{eq:12}) shows that crossover in the ground state of the term is possible at the same crystal field $10Dq=3J_H$ as for the $3d^5$ configuration.
The energy of ground states in Eqs.(\ref{eq:10}) and (\ref{eq:12}) in the pressure range $P_{min}<P<P_S$ promote the forbidden $fes$ quasiparticles (see Fig.\ref{fig:2}), and the $n$ - doped iron borate, can turn out to be an insulator (semiconductor), where
\begin{equation}
\chi _{r_0r_0}^\sigma  = {\sum\limits_{\lambda}{\gamma _{\lambda \sigma }^* \left( {{}^6A_{1} ,{}^5E } \right)\gamma _{\lambda \sigma } \left( {{}^5E ,{}^6A_{1} } \right)}}=0
\label{eq:121}
\end{equation}
at $\delta \left( {S'=0,S=5/2 \pm |\sigma| } \right)\delta \left( {M', M + \sigma } \right)=0$.
The disappearance of $n$ quasiparticles precedes the spin crossover at the pressure $P_S$. Note, however, the iron borate has so far not been able to be converted into the metallic state with increasing pressure.

\subsection{\label{sec:PIE}  Quasiparticles induced by pseudo JT effect\\}

\begin{figure*}
\includegraphics{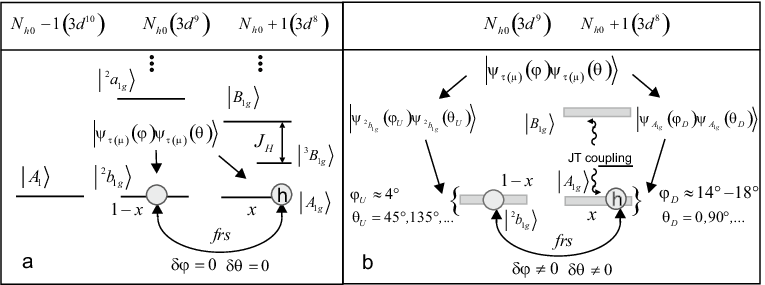}
\caption{\label{fig:3} $E_i({N_{h}},{M_S})$ - energy level scheme of the configuration space based on the cell states (\ref{eq:14}) in CuO$_2$ layer.  The solid line with arrows corresponds to the  $frs$ quasiparticles. (a) The tilting and orientation angles $\varphi$ and $\theta$ of the CuO$_6$ octahedron are saved by the $frs$ quasiparticles. (b) The wavy line shows the states active in the pseudo JT effect. The angles do not have specific magnitudes and are changed by $frs$ quasiparticles}
\label{fig:3}
\end{figure*}

A second possibility to observe the effects of low quasiparticle transparency is the the material with a dynamic JT effect, where the doped hole changes the initial orbital configuration of $\left| ({N_{h0},{M_{S}}})_{\mu=0} \right\rangle$ ground cell states ($l$ - forbidden $frs$ quasiparticles). Indeed single-particle operators $c_{f\lambda\sigma}^{(+)}$ in the matrix elements (\ref{eq:2}) and quasiparticle transparence (\ref{eq:3}) are non-diagonal operators, and quenching effects can be expected.

As follows from the works~\cite{Grissonnanche2020, Grissonnanche2019} the phonons have zero termal Hall response outside the pseudogap phase in the hole doped 2D perovskite cuprates. However, inside the pseudogap phase, the phonons become chiral to generate the Hall response.  It's show a specific symmetrical nature of the electron-lattice coupling. Next we apply the MI criterion, to understand the symmetric nature of phonons in the pseudogap effect of the hole spectrum in doped 2D perovskite cuprates. At first, we will check the MI criterion for doped cuprates in the frameworks of usual Russell-Saunders scheme. Let's calculate the magnitude of $N^{s(t)}_{frs}$ in the cuprates, where $r=\{^2b_1,A_{1g}\}$ - root vector are relevant in Eq.(\ref{eq:A.9}) at $\mu=0$ and $\tau=0$,~\cite{Feiner_etal1996, Gavrichkov_etal2001} i.e. it corresponds to only the $A_{1g}$ singlet $frs$ state. It's the well known Zhang-Rice singlet~\cite{Zhang1988}. Using exact diagonalization procedure with LDA parameters from the work~\cite{Korshunov_etal2005}, we obtain the relation:
\begin{eqnarray}
N_{frs}^s&=&1+[\beta^2_0(h_b)-\beta^2_0(h_{d_x})]\times \nonumber \\
&\times&[B^2_0(h^2_b)-B^2_0(h^2_{d_x})]\approx0.97
\label{eq:13}
\end{eqnarray}
for the singlet $frs$ states, where the doublet and singlet ground states (\ref{eq:5}) and (\ref{eq:6}) are
\begin{eqnarray}
|^2b_1\rangle_0&=&\beta_0(h_b)|h_b,\sigma_{\frac{1}{2}}\rangle+\beta_0(h_{d_x})|h_{d_x},\sigma_{\frac{1}{2}}\rangle \nonumber \\
|A_{1g}\rangle_0&=&B_0(h^2_b)|h^2_b,0_0\rangle+B_0(h^2_{d_x})|h^2_{d_x},0_0\rangle+ \nonumber \\
&+&B_0(h_{d_x},h_b)|h_{d_x},h_b,0_0\rangle,
\label{eq:14}
\end{eqnarray}
$h_b$ and $h_{d_x}$ - holes in the $b$-symmetrized cell states of oxygen and $d_{x^2-y^2}$ cooper states of CuO$_2$ layer respectively. Thus, there is no any $l$ - forbidden quasiparticles in the Russell-Saunders scheme, and the number of valence states is practically a constant: $N_v(x)\approx N_e-0.03x$. Here, the $frs$ quasiparticles are associated with single-hole transitions in Fig.\ref{fig:3}a, where the lattice of CuO$_2$ layer  is unchanged, and adiabatic approximation is correct.

In the dynamic JT effect, the CuO$_6$ octahedra can be in both the U stripe, and the D stripe states with different tilting and orientation angles $\varphi_{U(D)}$ and $\theta_{U(D)}$ respectively ~\cite{Gavrichkov2019}. The hole concentration $x$ in them is also different, and the dynamic JT effect in the CuO$_2$ layer as a whole would be possible only if its total charge and regular stripe structure were saved. Let's assume that the tilts $\varphi_{U(D)}$ and orientations $\theta_{U(D)}$ of the CuO$_6$  octahedra with respect to the spacer rock salt layers are the active JT distortions, and a rotation of all tilted CuO$_6$ octahedra around the $c$ axis (i.e. changing their orientation $\theta$) in a stripe U/D/U/D... structure fits within these limitations ~\cite{Gavrichkov2022}. However, a scale of the novel JT cell in the stripe set U/D/U/D... exceeds the initial cell (i.e. the single CuO$_6$ octahedron). As a result, in the dynamic JT effect, the hole number $x$ are not saved in the single CuO$_6$ octahedron under rotation around the $c$ axis.

In Eqs.(\ref{eq:2}) and (\ref{eq:3}), where matrix elements $\gamma_{\lambda\sigma}(r)$ were calculated still in local Russell-Saunders scheme, the overlapping phonon parts of the initial cell functions (\ref{eq:A.1}) and (\ref{eq:A.2}) arises. In meaning, it's Ham's reduction factor ~\cite{Ham1965} in the quasiparticle transparence, because the JT cell has a fourfold degeneracy  with different $\theta_{U(D)}$ orientations of tilted CuO$_6$ octahedra:

\begin{widetext}
\begin{equation}
\chi _{r_0r_0}^\sigma (\delta\theta, \delta\varphi)  = {\gamma _{x^2 \sigma }^* \left( {{}^1A_{1g} ,{}^2b_{1g} } \right)\gamma _{x^2 \sigma } \left( {{}^2b_{1g}{},^1A_{1g} } \right)} \cdot {\sum\limits_{\delta\varphi\delta\theta}\alpha(\delta\varphi,\delta\theta)},
\label{eq:15}
\end{equation}
\end{widetext}
where $r_0=\{^1A_{1g} ,^2b_{1g}\}$, $\theta=\theta_U, \theta_D$ and $\varphi=\varphi_U, \varphi_D$ with the indices U  and D  related to the stripe affiliation for the single CuO$_6$ octahedron in the harmonic oscillator states   $\left| {\psi _{\tau(\mu)}(\theta) } \right\rangle$ and $\left| {\psi _{\tau(\mu)}(\varphi) } \right\rangle$ of the displaced $2D$ oscillator~\cite{Gavrichkov2022}:
\begin{widetext}
\begin{equation}
\alpha(\delta\varphi,\delta\theta)=\left\langle {\psi _{^1A_{1g}}(\theta _D)^{} } \right|\left. {\psi _{^2b_{1g}}(\theta _U )^{} } \right\rangle^2  \cdot \left\langle {\psi _{^1A_{1g}}(\varphi_D)^{} } \right|\left. {\psi _ {^2b_{1g} }(\varphi_U )^{} } \right\rangle^2 \approx \exp \left\{ { - \nu{{\left( {\delta \theta _{}^2  + \delta \varphi_{}^2 } \right)} \mathord{\left/
{\vphantom {{\left( {\delta \theta _{}^2  + \delta \alpha _{}^2 } \right)} 2}} \right.
\kern-\nulldelimiterspace} 2}} \right\}
\label{eq:16}
\end{equation}
\end{widetext}
where $\delta \varphi  = \varphi_D  - \varphi_U  \approx 9^\circ - 13^\circ$, $\delta \theta  = \theta_D  - \theta_U  = \pm45^\circ$ and  $\nu=K/\hbar\omega_D$ with a Debye frequency $\omega_D$ and force coefficient $K$. Formally, the hole-doped 2D cuprate becomes an insulator only at $\omega_D\rightarrow0$. We will obtain the reduction effects in Eq.(\ref{eq:16}), if the charge inhomogeneity of the dynamical U/D/U/D... stripe structure (see Fig.(\ref{fig:3}b) occurs. To illustrate the last conclusion we considered a simplified model in the Appendix~\ref{ap:2}.


\section{\label{sec:IV} Conclusions\\}

In accordance with Wilson's MI criterion for the materials with strong electronic correlations, a class of materials with the forbidden $frs(fes)$ quasiparticles  in the Hubbard gap was identified. In order to highlight the nature of the specific materials, we introduced the term "quasiparticle transparency" for the matrix $\chi _{rr'}^\sigma$ (see Eq.(\ref{eq:3})), which in the materials can have a zero magnitude, and the doping does not lead to the charge carriers generation. Actually, with the term we note the analogy of the $frs(fes)$ quasiparticles with a propagation of light in a material with optical intracell absorption at frequencies of the $dd$ transitions.

A first tentative search for materials with the low quasiparticle transparency allowed us to identify the following materials as candidates:

(i) the $3d$ oxides with a spin crossover (e.g. FeBO$_3$ and other transition element oxides: LaMnO$_3$($3d^4$), Fe$_2$O$_3$($3d^5$), MnO($3d^5$), LaCoO$_3$($3d^6$), CoO($3d^7$), LaCoO$_4$($3d^7$), LaNiO$_3$($3d^7$) ..., where the low quasiparticle transparency effects can be observed in the vicinity of the crossover at high pressure (e.g. $P_S=48 - 54$ GPa for the iron borate). The doped charge carriers are generated and disappear with increasing pressure.

(ii) the materials with dynamic JT effect (2D perovskite cuprates), where in the  hole-doped materials the overlapping of phonon functions in the different sectors $N_{h0}$ and $N_{h0}+1$ of cell configuration space leads to partial quenching of charge carriers. An essence of the quenching is that even in the dynamic JT effect, the doped holes avoid the U stripes, and the threshold nature of the pseudo Jahn Teller  effect  in the doping concentration $x_D>x_c$ in the D stripes continues to support the charge inhomogeneous U/D/U/D... stripe structure. Otherwise, the hole concentration drops below critical $x<x_c$ and the pseudo JT effect disappears~\cite{Gavrichkov2024}. However, we are forced to pay for the dynamic stripe structure by introducing the bifurcation nature of the adiabatic potential in Fig.\ref{fig:3}b.

The real question is how can the doped materials with low quasiparticle transparency be identified? In undoped materials, the signatures of forbidden $frs(fes)$ quasiparticles in the single-particle density of states are missing. However, they can be enhanced by a resonant optical excitation, since a forrbiden on the nonzero magnitudes (\ref{eq:2}) does not apply to the optical matrix elements. Therefore, it is possible to detect the low quasiparticle transparency  by studying a difference between the optical gap and photoconductivity measurements for a mobility gap(see e.g. the work~\cite{Yu1992}). The first corresponds to a charge transfer gap (so called a CT gap) in the cuprates, where light-induced $frs$ "quasiparticles" are localized, and the second corresponds to a gap in the spectrum of charge carriers. As follows from our work, the difference in the undoped cuprates is directly related to the pseudogap effect in the hole-doped cuprates. The photoconductivity has been mainly a topic of superconductor research during the 1990s, the explanation for the effect is still under some debate~\cite{Destraz2015}.\\

\begin{acknowledgments}
We acknowledge the support of the Russian Science Foundation through grant RSF No.22-22-00298.
\end{acknowledgments}

\appendix
\section{\label{ap:1} The number of frs(fes) states in doped Mott-Hubbard materials}
In a case of one hole per cell, ${\left| {({N_{h}},{M_S})_i} \right\rangle} $ cell states are a superposition of different hole configurations of the same orbital ($l$ -)symmetry:
\begin{equation}
\left| ({N_{h0},{M_{S}}})_{\mu} \right\rangle  = \sum\limits_\lambda ^{} {{\beta _\mu }\left( {{h_\lambda }} \right)\left| {{h_\lambda },{M_{S}}} \right\rangle } |\psi_\mu(\varphi)\psi_{\mu}(\theta)\rangle
\label{eq:A.1}
\end{equation}
Thus, there are  $C_{{2N_\lambda}}^{1} = {2N_\lambda}$ the one-hole spin doublet states, where  $C^k_n$ -  number of combinations. Altogether there are  $C^2_{2N_\lambda}=N_S+3N_T$ of the spin singlets $N_S=C_{N_\lambda}^{2}+N_\lambda$ and triplets $N_T=C_{N_\lambda}^{2}$:
\begin{eqnarray}
\left| ({N_{h0}+1,{M'_{S'}}})_{\tau} \right\rangle  =&& \sum\limits_{\nu\nu'} {{B_\tau }\left( {{h_\nu},{h_{\nu'}}} \right)\left| {{h_\nu},{h_{\nu'}},{M'_{S'}}} \right\rangle }\times \nonumber \\
&&\times|\psi_{\tau}(\varphi)\psi_{\tau}(\theta)\rangle
\label{eq:A.2}
\end{eqnarray}
in the two-hole sector (see Fig.\ref{fig:1}) of the ${N_\lambda}$ - orbital approach, where the harmonic oscillator wave function   $\left| \psi_{\mu(\tau)}(\varphi) \psi_{(\mu(\tau)}(\theta) \right\rangle=\left| \psi_{\mu(\tau)}(\varphi) \right\rangle\left| \psi_{\mu(\tau)}(\theta)  \right\rangle$ associated with the (non-)displaced $2D$ oscillator~\cite{Gavrichkov2022}. Using the intracell Hamiltonian $H_0$
in the cell functions  representation
the configuration weights $\beta_\mu(h_\lambda)$ and $B_\tau(h_\lambda,h_{\lambda'})$ can be obtained by the exact diagonalization procedure  for the matrix $(\hat{H_0})_{\lambda\lambda'}$ and $(\hat{H_0})^{\nu\nu'}_{\lambda\lambda'}$ in the $E_i({N_{h},{M_S}})$-eigenvalue problem  at the different sectors $N_h$.~\cite{Ovchinnikov_etal2012}

A sum (\ref{eq:5}) over all the $r$-th excited states with $\mu\neq0$ in the  sector $N_h=N_{h0}$ is omitted, and only the excited states with any $\tau$($\eta$) index in the  nearest $N_h=(N_{h0}+1)$ and $(N_{h0}+2)$ sectors are summed up. The expressions for high- and low-spin two-hole partner states (with $S'=S\pm|\sigma|$) can be combined into a single expression:

\begin{widetext}
\begin{equation}
\left| {h_{\lambda},{h_{\lambda '} ,M'_{S'}}} \right\rangle  = \{\Gamma _ \uparrow ^{}\left( {S'_{M'},S} \right)c_{\lambda ' \downarrow } \left| {{h_\lambda },{M'-\textstyle{1 \over 2}}} \right\rangle  + {\mathop{\rm sgn}} (\Delta S)\Gamma _ \downarrow ^{}\left( {S'_{M'},S} \right)c_{\lambda ' \uparrow} \left| {{h_\lambda },{M'+\textstyle{1 \over 2}}} \right\rangle\}
\label{eq:A.3}
\end{equation}
\end{widetext}
where $\Delta S=S'-S=\pm|\sigma|$, and the coefficients
\begin{equation}
\Gamma _\sigma ^2\left( {S'_{M'},S} \right) = \frac{{S + \eta \left( \sigma  \right){\mathop{\rm sgn}} \left( {\Delta S} \right)M' + {\textstyle{1 \over 2}}}}{{2S + 1}}
\label{eq:A.4}
\end{equation}
have the completenessn property for contributions from the identical spin states of doped hole to the different high- and low-spin two-hole partners:
\begin{equation}
\sum\limits_{\Delta S = - \left| \sigma  \right|}^{ + \left| \sigma  \right|} {\Gamma _\sigma ^2\left( {S'_{M'},S } \right)} = \sum\limits_{\sigma}^{ }{\Gamma _\sigma ^2\left( {S'_{M'},S } \right) = 1},
\label{eq:A.5}
\end{equation}
and also
\begin{equation}
\sum\limits_{M =  - S}^S {{\Gamma _\sigma ^2\left( {{{S'}_{M'}},S} \right)} }
=S + \frac{1}{2}
\label{eq:A.6}
\end{equation}
 Taking into account relations (\ref{eq:A.1}), (\ref{eq:A.2}) and (\ref{eq:A.6}) we can determine the matrix element in Eq.(\ref{eq:5}) by the sum:
\begin{widetext}
\begin{eqnarray}
\left\langle ({{N_{h0}} + 1,{{M'}_{S'}}})_\tau \right|c_{\nu \sigma } \left| ({{N_{h0}},{M_S}})_\mu \right\rangle = &&\sum\limits_{\lambda ,\lambda ',\lambda ''} {\left\langle {{h_{\lambda '}},{h_{\lambda ''}},M'_{S'}\left| {c_{\nu \sigma }} \right|{h_\lambda },M_{S}} \right\rangle {\beta _\mu }\left( {{h_\lambda }} \right){B_\tau }\left( {{h_{\lambda '}},{h_{\lambda ''}}} \right)\Gamma _\sigma \left( {S'_{M'} ,S} \right)}\times \nonumber \\
&&\times\left\langle {\psi _{\tau}(\varphi) \psi _{\tau}(\theta)} \right|\left. {\psi _{\mu}(\varphi)\psi_{\mu}(\theta)} \right\rangle.
\label{eq:A.7}
\end{eqnarray}
\end{widetext}
After substituting Green's function (\ref{eq:1}), the expressions (\ref{eq:2}) and (\ref{eq:A.6})  to the Eq.(\ref{eq:5}) we obtain:
\begin{equation}
N_v (x)= N_v^{12}+N_v^{23}={N^{12}_{s,v}} + 3{N^{12}_{t,v}}+N_v^{23},
\label{eq:A.8}
\end{equation}
where instead of the sum over root vectors ${{r}}$, we used the summation over a physically meaningful indices $\tau $,  $M$ and $\Delta S$
(i.e. the sum over all low ($s$) - and high-spin ($t$) two-hole states). Here is

\begin{widetext}
\begin{eqnarray}
N_{s(t),v}^{12}&=& \sum\limits_{\nu \sigma } {\sum\limits_{\tau }
{{F^{s(t)}_{r = \left\{ {0,\tau } \right\}}}\left( x \right) \sum\limits_{MM'}^{}
{{{\left\{ {\sum\limits_\lambda  {\Gamma _\sigma \left( {S'_{M'}, S} \right){\beta _{\mu  = 0}}
\left( {{h_\lambda }} \right)} {B_\tau }\left( {{h_\lambda },{h_\nu }} \right)\delta \left( {S',S \pm |\sigma| } \right)
\delta \left( {M', M +  \sigma} \right)} \right\}}^2}} } }\times \nonumber \\
&&\times {\sum\limits_{\varphi \theta } \left\langle {\psi _{\tau}(\varphi) \psi _{\tau}(\theta)} \right|\left. {\psi _{\mu=0}(\varphi)\psi_{\mu=0}(\theta)} \right\rangle^2},
\label{eq:A.9}
\end{eqnarray}
\end{widetext}
where the $(+)$ and $(-)$ on the right side are used with indices $t$ and $s$ respectively, occupation factor in the PM phase:
\begin{equation}
{F^{s(t)}_{\left\{ {0,\tau } \right\}}}(x) = \left\{ {\begin{array}{*{20}{c}}
   {\tfrac{1}{2}\left( {1 - \alpha_{s(t)} x} \right),\tau  = 0}  \\
   {\tfrac{1}{2}\left( {1 - x} \right),\tau  \ne 0}  \\
\end{array} } \right.,
 \label{eq:A.10}
\end{equation}
 with $\alpha_{s(t)}=1-{2}/{(2S'+1)}$ and $S'=0,1$; $S=1/2$. Let's start with the contribution from the spin singlet $frs$ states $N^s_{frs}$:

\begin{equation}
N_{v}(x) =(2N_\lambda-1)-x(1-N^s_{frs})=N_e-x(1-N^s_{frs}),
\label{eq:A.11}
\end{equation}
where the low and high spin contributions are
\begin{equation}
N^{12}_{s,v}=(1/2)[(N_\lambda+1)(1-x)+2xN^{s}_{frs}]
\label{eq:A.12}
\end{equation}
 and
\begin{equation}
N^{12}_{t,v}=(1/2)(N_\lambda-1)(1-x)
\label{eq:A.13}
\end{equation}
respectively. In the static case,

\begin{eqnarray}
\sum\limits_{\varphi \theta } \left\langle {\psi _{\tau}(\varphi) \psi _{\tau}(\theta)} \right|\left. {\psi _{\mu}(\varphi)\psi_{\mu}(\theta)} \right\rangle^2&&= \label{eq:131} \\
=\left\langle {\psi _{\tau}(\varphi_{U,D}) \psi _{\tau}(\theta_{U,D})} \right|&&\left. {\psi _{\mu}(\varphi_{U,D})\psi_{\mu}(\theta_{U,D})} \right\rangle^2=1,
\nonumber
\end{eqnarray}
at the any $|\tau\rangle$ and $|\mu\rangle$ electron states because the tilting $\varphi=\varphi_{U,D}$ and orientation $\theta=\theta_{U,D}$ angles are fixed and associated with the minimum of the single adiabatic potential in the U and D stripes. The relation $N_v^{23}=x(2N_\lambda-2)$ for the contributions from the quasiparticle with root vectors from $\{r_{23}\}$ subspace is derived similarly to the previous expression for the contribution (\ref{eq:A.9}).
The number of possible singlet $frs$ states is in the range $0 \leq N^s_{frs} \leq 2$, where
\begin{eqnarray}
N_{frs}^s &=& 1 - \sum\limits_\lambda  {\beta _0^2\left( {{h_\lambda }} \right)}\times\nonumber\\ &&\sum\limits_{\lambda ',\lambda ''} {\left[ {1 - {\delta _{\lambda \lambda '}} - {\delta _{\lambda \lambda "}}} \right]B_{\tau  = 0}^2\left( {{h_{\lambda '}}{h_{\lambda ''}}} \right)},
\label{eq:A.14}
\end{eqnarray}
and $ \tau  = 0   $ corresponds to the $frs$-quasiparticles. In deriving Eq.(\ref{eq:A.11}) we also used the Eq.(\ref{eq:A.6}) and identity $\sum\limits_\lambda  {\beta _\mu ^2\left( {{h_\lambda }} \right)} \sum\limits_{\lambda ',\lambda ''} {B_\tau ^2\left( {{h_{\lambda '}}{h_{\lambda ''}}} \right)}  = 1$ at any $\mu$  and $\tau$. Because the sum
\begin{equation}
\sum\limits_\tau  {\left[ {{\beta _0}\left( {{h_\lambda }} \right){B_\tau }\left( {{h_\lambda },{h_\nu }} \right)} \right]\left[ {{\beta _0}\left( {{h_{\lambda '}}} \right){B_\tau }\left( {{h_{\lambda '}},{h_\nu }} \right)} \right]}  = 0
\label{eq:A.15}
\end{equation}
at the any $\nu $ and $\lambda  \ne \lambda '$, the contribution from cross-term  from (\ref{eq:A.9}) to the total number of the valence states is missing.
In a case of triplet nature of the $frs$ states, we obtain a similar expression to (\ref{eq:A.11}) with the
contribution
\begin{equation}
N_{frs}^t = 1 - \sum\limits_\lambda  {\beta _0^2\left( {{h_\lambda }} \right)} \sum\limits_{\lambda ' \ne \lambda '' \ne \lambda } {B_{\tau  = 0}^2\left( {{h_{\lambda '}}{h_{\lambda ''}}} \right)},
\label{eq:A.16}
\end{equation}
where $0 \le {N_{frs}^t} \le 1$.\\

\section{\label{ap:2} The number of frs states in doped Mott-Hubbard materials with pseudo JT effect}
To show a role of the charge inhomogeneity, let's obtain the MI criterion in a simpler three-orbital model ($\lambda  = a,b,c$) of a semiconductor with two spinless electrons and hole doping $N_e=2-x$.
Here are $\left| \mu  \right\rangle  = a_\mu  \left| a \right\rangle  + b_\mu  \left| b \right\rangle  + c_\mu  \left| c \right\rangle$
and $\left| \tau  \right\rangle  = \left( {ab} \right)_\tau  \left| a \right\rangle  + \left( {bc} \right)_\tau  \left| {bc} \right\rangle  + \left( {ac} \right)_\tau  \left| {ac} \right\rangle$, where $\left| \mu  \right\rangle$ and $\left| \tau \right\rangle$  are the states with $\mu \left( \tau  \right) = 0 - 2$ in the sectors $N_{h0}$  and $N_{h0}+1$  respectively.
The coefficients satisfy the completeness relations $a_\mu ^2  + b_\mu ^2  + c_\mu ^2  = 1$  and  .  $\left( {ab} \right)_\tau ^2  + \left( {bc} \right)_\tau ^2  + \left( {ac} \right)_\tau ^2  = 1$.
In the homogeneous undoped case $\left( {x = 0} \right)$, the number of valence states
\begin{equation}
N_v  = N_v^{12}  = \sum\limits_\tau  {\left\{ {1 - a^2\left( {bc} \right)_\tau ^2 - b^2\left( {ac} \right)_\tau ^2 - c^2\left( {ab} \right)_\tau ^2 } \right\}}  = 2
\label{eq:B.1}
\end{equation}
since $\sum\limits_\tau  {\left( {bc} \right)_\tau ^2 }  =\sum\limits_\tau  {\left( {ab} \right)_\tau ^2 }=\sum\limits_\tau  {\left( {ac} \right)_\tau ^2 }= 1$  the material, according to the criterion, is a semiconductor, at $N_e  = 2$. Let's now, $x \ne 0$ ,  then
\begin{widetext}
\begin{equation}
N_v \left( x \right) = N_v^{12}  + N_v^{23}  = N_{frs} + \left( {1 - x} \right)\sum\limits_{\tau=1}^2\left\{ 1 - a^2_0\left( {bc} \right)_\tau^2 - b^2_0\left( {ac} \right)_\tau^2-c^2_0\left( {ab} \right)_\tau^2 \right\} + N_v^{23}  = 2 - \left\{ {1 - N_{frs} } \right\}x,
\label{eq:B.2}
\end{equation}
\end{widetext}
where $N_{frs}  = 1 - a^2_0\left( {bc} \right)_0^2-b^2_0\left( {ac} \right)_0^2-c^2_0\left( {ac} \right)_0^2$, and $N_v^{23}  = x$.

Now let us introduce into this model the JT instability in the two-particle sector, as is assumed for cuprates ~\cite{Gavrichkov2022}, where the JT distortions are the angles $\varphi$ and $\theta$  of the tilting and orientation of the CuO$_6$ octahedra. In the dynamic state, all octahedra do not have specific tilts $\varphi$ and orientations $\theta$, and all states $\left| {\tau \left( \mu  \right)} \right\rangle \left| {\psi _{\mu \left( \tau  \right)} \left( \varphi  \right)\psi _{\mu \left( \tau  \right)} \left( \theta  \right)} \right\rangle$ cannot be occupied with the factors $\left( {1 - x} \right)$ and $x$ in both sectors $N_{h0}$ and $N_{h0}+1$ of the configuration space. Indeed, in this case the charge inhomogeneity disappears along with the pseudo JT effect at the concentration  below the threshold $x<x_c$. We can choose the bifurcation potential in Fig.\ref{fig:3}b, at which $x$ doped carriers still avoid U stripes, so that the number of valence states at $x \ne 0$   is
\begin{equation}
N_v \left( x \right) \approx 2 - \left\{ {1 - N_{frs} } \right\}x \cdot \alpha \left( {\delta \varphi ,\delta \theta } \right),
\label{eq:B.3}
\end{equation}
where 
\begin{widetext}
\begin{equation}
\alpha \left( {\delta \varphi ,\delta \theta } \right) = {\sum\limits_{\varphi _D \theta _D } {\sum\limits_{\varphi _U \theta _U } {\left\langle {\psi _{\mu  = 0} \left( {\varphi _D } \right)\psi _{\mu  = 0} \left( {\theta _D } \right)\left| {\psi _{\tau  = 0} \left( {\varphi _U } \right)\psi _{\tau  = 0} \left( {\theta _U } \right)} \right.} \right\rangle ^2 } } } \le 1,
\label{eq:B.4}
\end{equation}
\end{widetext}
and the number of valence states $N_v \left( x \right)$ decreases. Depending on the magnitude of $\alpha \left( {\delta \varphi ,\delta \theta } \right)$, the criterion can detect the ground state of doped JT material close to insulating.

\bibliography{my}
\end{document}